\documentclass[aps,prl,twocolumn,showpacs,superscriptaddress,groupedaddress]{revtex4}  
\usepackage{graphicx}  
\usepackage{dcolumn}   
\usepackage{bm}        
\usepackage{amssymb}   
\usepackage{amsmath}

\newcommand{\be}{\begin{equation}}
\newcommand{\ee}{\end{equation}}

\newcommand{\bea}{\begin{eqnarray}}
\newcommand{\eea}{\end{eqnarray}}

\begin{document}

\hspace{5.2in} \mbox{IGC-12/12-3}

\title{Statistical Naturalness and non-Gaussianity in a Finite Universe}
\author{Elliot Nelson\footnote{eln121@psu.edu}$^1$, Sarah Shandera\footnote{shandera@gravity.psu.edu}}
\affiliation{Institute for Gravitation and the Cosmos, The Pennsylvania State University, University Park PA 16802}

\begin{abstract}

We study the behavior of $n$-point functions of the primordial curvature perturbations, assuming
our observed universe is only a subset of a larger space with statistically homogeneous and isotropic
perturbations. If the larger space has arbitrary n-point functions in a family of local type non-
Gaussian statistics, sufficiently biased smaller volumes will have statistics from a `natural' version
of that family with moments that are weakly non-Gaussian and ordered, regardless of the statistics
of the original field. We also describe the effect of this bias on the shape of the bispectrum.

\end{abstract}

\maketitle

Measurements of the primordial density fluctuations are the primary tool to test the paradigm of inflationary cosmology and to distinguish between the many proposed particle physics scenarios for inflation. As our ability to test the statistics beyond the power spectrum, collectively called non-Gaussianity, becomes more advanced, new questions arise: what is the best way to test non-Gaussianity? What measurements would point definitively to particular models of inflation? So far the proposed approaches to address these questions rely either on a particle physics notion of naturalness for non-Gaussianity, either for the inflaton field \cite{Weinberg:2008hq} or for the fluctuations \cite{Cheung:2007st}, or on mode expansions to try to capture any non-Gaussianity that is observationally accessible \cite{Fergusson:2009nv}.

Here we point out a distinct and complementary way of thinking about naturalness. We suppose only that the universe is considerably larger than what we see (which is the natural outcome in many inflation models) and that there exists a homogeneous and isotropic spectrum of primordial fluctuations in the gravitational field on all scales in the entire volume. If the field is non-Gaussian, statistics in any given spatial subset may be biased in comparison to the global statistics due to coupling of modes in the subset to long-wavelength background modes. The relevance of this effect increases as the subvolume size decreases because there are more long-wavelength modes. In smaller subvolumes, local statistics are typically more biased, and vary more from region to region. In the case of exactly Gaussian statistics, there is no coupling to long-wavelength modes and the only effect of biasing is to shift the locally determined mean of the fluctuations.

For a given choice of statistics in the large volume, we can ask what statistics are typical to spatial subsets of the size of our universe. In this report, we will find a notion of statistical naturalness for typical small volumes where a family of well-behaved correlation functions is generated from a parent volume with arbitrarily fine-tuned statistics in the same family. Our results build on previous work on the non-Gaussian halo bias from the standard local ansatz \cite{Dalal:2007cu} and in $g_{NL}$ type non-Gaussianity \cite{Smith:2011ub}, and can be used to more precisely characterize observable features of multi-field inflation models \cite{JasonJoeUs}. This work is an extension of ideas in \cite{Boubekeur:2005fj} and is similar in spirit to recent results in \cite{Byrnes:2011ri, Schmidt:2012ky,Tasinato:2012js,Nurmi:2013xv}. Related work on the effects of superhorizon fluctuations in the context of large scale anomalies in the cosmic microwave background includes \cite{Gordon:2005ai,Erickcek:2008sm}. We illustrate our point with a simple first example, showing that the local ansatz for non-Gaussianity, with an amplitude that is weakly non-Gaussian and whose principal term is quadratic in the underlying Gaussian, is statistically natural.

Consider a large volume characterized by side length or radius $L$ and a smaller volume characterized by scale $M$. Here we will generally have in mind that $M$ is the scale of our currently observable universe and $L$ the scale of the entire universe (which we assume to be finite). Note that $L$ may also just be the largest scale on which this prescription for the fluctuations is trusted. It is also sometimes useful to consider $L$ to be the size of our observable universe and $M$ the scale of an N-body simulation or of some local region whose Large Scale Structure we are interested in. We define the curvature perturbation in each region as the fractional shift to the scale factor $a$ describing a background, homogeneous Friedmann-Robertson-Walker universe:
\bea
a(x)&=&\bar{a}_L(1+\tilde{\zeta}(x))\;,\;\;\;x\in Vol_L\\
&=&\bar{a}_M(1+\zeta(x))\;,\;\;\;x\in Vol_M
\eea
where $|\tilde{\zeta}|$, $|\zeta|<1$ by definition. We define a maximum wavenumber $k_{\text{max}}$ from the smallest scale we smooth over in defining the fluctuations. In the second line, we have simply defined $\zeta$ in the subvolume $Vol_M$ as the fluctuations around the \textit{local} background $\bar{a}_M=\bar{a}_L(1+\langle\tilde{\zeta}\rangle_M)$, where $\langle\tilde{\zeta}\rangle_M$ is the average over $Vol_M$ of the perturbations defined with respect to the volume $L$. These quantities are related by
\be
1+\tilde{\zeta}(x)=(1+\langle\tilde{\zeta}\rangle_M)(1+\zeta(x))\;,\;\;\; x\in Vol_M. \label{zetaLM}
\ee
The power spectrum in either volume is defined in terms of the two-point function, $\langle\zeta_{\mathbf{k}_1}\zeta_{\mathbf{k}_2}\rangle\equiv(2\pi)^3\delta^3(\mathbf{k}_1+\mathbf{k}_2)P(k)$, and the dimensionless power spectrum is $\mathcal{P}(k)\equiv\frac{k^3}{2\pi^2}P(k)$.

\textit{Local Ansatz.}
Consider a simple form of the local ansatz where the curvature perturbation, $\tilde{\zeta}(x)$ is a local non-linear function of a Gaussian field $\tilde{\zeta}_G(x)$. In the large volume, suppose the curvature perturbation is
\be
\tilde{\zeta}(x)=\tilde{N}_1\tilde{\zeta}_G(x)+\frac{1}{2!}\tilde{N}_2\tilde{\zeta}_G(x)^2+\frac{1}{3!}\tilde{N}_3\tilde{\zeta}_G(x)^3+..., \label{zetaL2}
\ee
where we implicitly shift the field so the mean $\langle\tilde{\zeta}\rangle$ throughout the large volume is zero. The original local ansatz \cite{Salopek:1990jq} set $\tilde{N}_1=1$ and $\tilde{N}_2=\frac{6}{5}\tilde{f}_{NL}$ to define the non-linearity parameter $\tilde{f}_{NL}$. We will take the $\tilde{N}_i$ to be constants.  The real-space variance of the Gaussian field $\tilde{\zeta}_G$ is
\be
\tilde{\sigma}_0^2\equiv\langle\tilde{\zeta}_G^2\rangle=\int_{L^{-1}}^{k_{\text{max}}}\frac{d^3k}{(2\pi)^3}\tilde{P}_G(k),
\ee
where $\tilde{P}_G$ is the power spectrum of $\tilde{\zeta}_G$. It will also be useful to define $\tilde{\sigma}^2_{0l}$ and $\tilde{\sigma}^2_{0s}$, with limits of integration changed to ($L^{-1},M^{-1}$) and ($M^{-1},k_{\text{max}}$), respectively.

Consider $\tilde\zeta(x)$ for $x\in M$, a subsample. Dividing $\tilde{\zeta}_{G}(x\in M)$ into long and short-wavelength parts gives $\tilde{\zeta}_M=\tilde{\zeta}_{l,M}+\tilde{\zeta}_{s,M}$. Here $\tilde{\zeta}_{l,M}$ is the real space field smoothed over region $M$ and is similar to $\langle\tilde{\zeta}\rangle_M$ defined above, up to the difference between the real space and Fourier space top-hat window functions. Following \eqref{zetaL2}, this gives the local background, a constant in any particular subsample $M$,
\be
\tilde{\zeta}_{l,M}=\tilde{N}_1\tilde{\sigma}_0B+\frac{1}{2!}\tilde{N}_2\tilde{\sigma}_0^2B^2+...,
\label{zetaLl}
\ee
where $B\equiv\tilde{\zeta}_{Gl,M}/\tilde{\sigma}_0$ is a measure of bias in a given subsample $M$. Similarly,
\be
\tilde{\zeta}_{s}(x)=\hat{N}_1\tilde{\zeta}_{Gs}(x)+\frac{1}{2!}\hat{N}_2\tilde{\zeta}_{Gs}^2(x)+\frac{1}{3!}\hat{N}_3\tilde{\zeta}_{Gs}^3(x)+... \;
\label{zetaLs2}
\ee
contains the short-wavelength fluctuations. However, the coefficients $\hat{N}_n$ now depend on the local background:
\be
\hat{N}_n(B)=\tilde{N}_n+\tilde{N}_{n+1}\tilde{\sigma}_0B+\frac{1}{2!}\tilde{N}_{n+2}\tilde{\sigma}_0^2B^2+... \label{N'n}
\ee
Then the curvature perturbation in any small volume,
\be
\zeta(x\in M)=\zeta_{G}(x)+\frac{1}{2!}N_2\zeta_{G}^2(x)+\frac{1}{3!}N_3\zeta_{G}^3(x)+..., \label{zetaM2}
\ee
is related to $\tilde{\zeta}$ by $\zeta=\tilde{\zeta}_s/(1+\tilde{\zeta}_{l,M})$, which follows from Eq. \eqref{zetaLM} and the long- and short-wavelength split above; here we have set $N_1\equiv1$. Reading off the Gaussian part of each expression, we have
\be
\zeta_{G}=\frac{\hat{N}_1}{1+\tilde{\zeta}_{l,M}}\tilde{\zeta}_{Gs}, \label{zetaGM2}
\ee
and consequently, the contribution $\mathcal{P}_{G}$ to the power spectrum from the Gaussian part $\zeta_G$ is different in the small volume, $\mathcal{P}_G=\left(\frac{\hat{N}_1}{1+\tilde{\zeta}_{l,M}}\right)^2\tilde{\mathcal{P}}_G$. We also define the real-space variance of $\zeta_G$,
\be
\sigma_0^2\equiv\langle\zeta_G^2\rangle=\int_{M^{-1}}^{k_{\text{max}}}\frac{d^3k}{(2\pi)^3}P_G(k).
\ee
The $N_n$ coefficients can be expressed in terms of the $\hat{N}_n$ coefficients,
\be
N_n(B)=\frac{(1+\tilde{\zeta}_{l,M}(B))^{n-1}}{\hat{N}^n_1(B)}\hat{N}_n(B), \label{Ntilde}
\ee
which can be verified by comparing Eq. \eqref{zetaLs2} with Eqs. \eqref{zetaM2} and \eqref{zetaGM2}. The $N_n$, and hence the locally averaged $n$-point functions, will vary among subsamples due to variation in the bias $B$, which is itself drawn from a Gaussian distribution with variance $\langle B^2\rangle=\tilde{\sigma}_{0l}^2/\tilde{\sigma}_0^2\leq1$. In the limit $M^{-1}\rightarrow k_{\text{max}}$, $\langle B^2\rangle\rightarrow1$.

The level of non-Gaussianity in $\tilde{\zeta}$ introduced by any one of the $\tilde{N}_n$ coefficients can be quantified by $\tilde{N}_n\tilde{\sigma}_0^{n-1}$.
Using Eqs. \eqref{zetaGM2} and \eqref{Ntilde}, it is easy to show that the corresponding quantity $N_n\sigma_0^{n-1}$ for the small volume is given by
\be
\lambda_n(B)\equiv N_n\sigma_0^{n-1}=\hat{N}^{-1}_1\hat{N}_n\tilde{\sigma}_{0s}^{n-1}. \label{NGtildeN}
\ee
The increase or decrease in the level of non-Gaussianity is determined by the same factor of $\hat{N}^{-1}_1$ for all terms, up to additional corrections in the $\hat{N}_n$, as expressed in Eq. \eqref{N'n}. If we truncate the series at two terms, where $\tilde{N}_2=\frac{6}{5}\tilde{f}_{NL}$, we find
\be
f_{NL}\sigma_0=\tilde{f}_{NL}\tilde{\sigma}_{0s}\left(1+\frac{6}{5}\tilde{f}_{NL}\tilde{\sigma}_0B\right)^{-1}. \label{fP}
\ee
Generically, if the series in the large volume $L$ was a good Taylor expansion with $\tilde{N}_{n+1}\tilde{\zeta}_{G}<\tilde{N}_n$, the coefficients in volume $M$ will be not too different from those in $L$. For unbiased subsamples, where the long wavelength modes happen to average to zero, $B=0$ and the statistics of the subsample are identical to those of the volume $L$.

The running of the parameters of the series with the background bias $B$ can also be expressed in differential form, analogous to renormalization group equations.  From Eq. \eqref{N'n} we have $\tilde{\sigma}_0^{-1}d\hat{N}_n/dB=\hat{N}_{n+1}$, and Eq. from \eqref{zetaLl} we have $\tilde{\sigma}_0^{-1}d\tilde{\zeta}_{l}/dB\equiv\tilde{\sigma}_0^{-1}d\hat{N}_0/d B=\hat{N}_1$.  One can show from Eq. \eqref{NGtildeN} that
\be
\frac{d\ln\lambda_n}{dB}=\frac{\lambda_{n+1}}{\lambda_n}-\lambda_2. \label{flow}
\ee
This equation is valid for any set of initial conditions ${\lambda_n(0)}$, that is, for any set of coefficients ${\tilde{N}_n}$, although one must take care when $B=0$ in cases where there is no linear term in the large volume ($\tilde{N}_1=0$), because in the small volume, Eq. \eqref{zetaM2}, we have normalized the linear term to have a coefficient $1$. Writing a similar differential equation for the dimensionless (connected) moments $\mathcal{M}_n\equiv\langle\zeta(x)^n\rangle_c/\langle\zeta(x)^2\rangle^{n/2}$ would avoid that problem and be more complete, but it is also more notationally cumbersome so we do not write it here.

\textit{Weakly Non-Gaussian Ansatz.} Let us now consider a case where the series in the volume $L$ is fine tuned, with some coefficients $\tilde{N}_n$ unusually large or small. Consider first the case where the $\tilde{N}_n$ with $p>n>1$ are zero in the large volume $L$ so that after the linear term the series starts only at order $p$:
\be
\tilde{\zeta}=\tilde{\zeta}_{G}+\frac{1}{p!}\tilde{N}_p\tilde{\zeta}_{G}^p+\frac{1}{(p+1)!}\tilde{N}_{p+1}\tilde{\zeta}_{G}^{p+1}+..., \ p\geq3. \label{WNG}
\ee
By ``weakly non-Gaussian'' we mean that the linear term dominates, so $\frac{1}{p!}\tilde{N}_p\tilde{\mathcal{P}}_G^{(p-1)/2}\ll1$. To ensure a simple behavior of the highest moments, we also assume that the nonzero terms become smaller by the same ratio $\tilde{r}\sim(\tilde{N}_{n+1}/\tilde{N}_n)\tilde{\mathcal{P}}_G^{1/2}\ll1$, and that $\frac{1}{p!}\tilde{N}_p\tilde{\mathcal{P}}_G^{(p-1)/2}\sim\tilde{r}^{p-1}$. This scenario gives a nearly Gaussian field in volume $L$ whose non-Gaussian moments have some unusual properties. For $n>p$ the dimensionless moments $\tilde{\mathcal{M}}_n$ scale like $\tilde{\mathcal{M}}_n\propto\tilde{r}^{n-2}$. However, the moments with $n\leq p+1$ are not necessarily ordered (eg, $\tilde{\mathcal{M}}_n\nless\tilde{\mathcal{M}}_{n+1}$ is possible). Defining $\tilde{A}\equiv\tilde{\sigma}_0^2/\tilde{\mathcal{P}}_G$, with $\tilde{A}\tilde{r}^2\ll1$, the moments with $n\leq p$ behave as $\tilde{\mathcal{M}}_n\propto\tilde{r}^p\tilde{A}^{\frac{p}{2}}$ for $(p,n)=(\text{odd},\text{odd})$ or $(\text{even},\text{even})$, and $\tilde{\mathcal{M}}_n\propto\tilde{r}^{p-1}\tilde{A}^{\frac{p-1}{2}}$ for $(p,n)=(\text{even},\text{odd})$ or $(\text{odd},\text{even})$.

However, in subsamples the long-wavelength modes will generate the missing lower order terms:
\be
\zeta=\zeta_{G}+\frac{1}{2!}N_2\zeta_{G}^2+\frac{1}{3!}N_3\zeta_{G}^3+...
\ee
With the restriction that the terms with $n>p$ in the large volume fall off according to $\tilde{r}\ll1$, $N_n\approx \tilde{N}_p(1+\tilde{\zeta}_{l}(B))^{n-1}(\tilde\sigma_0B)^{p-n}/(p-n)!$ for $p\geq n>1$, and $N_1\equiv1$. Interestingly, the correlation functions $\langle\zeta^n\rangle$ are not of order $N_2^{n-2}$ but are instead dominated by the contribution from $N_{n-1}$. 

For sufficiently biased subsamples, the series of dimensionless moments can be written, for $2<n\leq p$,
\be
\mathcal{M}_n
\propto\mathcal{C}[f_{NL}^{eff}\sigma_0]^{n-2}\;,\;\; \label{FTLmoments}
\ee
where $\mathcal{C}\propto\tilde{r}^{p-1}$, and $f_{NL}^{eff}=\frac{1+\tilde{\zeta}_{l}(B)}{\tilde\sigma_0B}$. That is, the level of non-Gaussianity and scaling of the moments in sufficiently biased subsamples is determined not by the original parameters $\tilde{N}_n$, but by the local background $B$. (In contrast, for $n>p$, $\mathcal{M}_n\propto \tilde{N}_{n-1}\sigma_0^{(n-2)/2}$.) 
Because $f_{NL}^{eff}\sigma_0\simeq\sigma_0/\tilde{\sigma}_0B$, we have $\mathcal{M}_{n+1}/\mathcal{M}_n\sim\frac{1}{B}\big(\frac{\ln(k_{\text{max}}M)}{\ln(k_{\text{max}}L)}\big)^{1/2}$. Consequently, for sufficiently biased subsamples, the $n\leq p$ moments will fall off as $n$ increases. We will see that this tends to be the case for subsamples containing fewer subhorizon modes than the number of superhorizon background modes. Note also that $\zeta$ is still only weakly non-Gaussian.

\textit{Strongly Non-Gaussian Ansatz.} Next, consider a case where the statistics in the volume $L$ are very non-Gaussian:
\be
\label{HNG}
\tilde{\zeta}=\frac{1}{p!}\tilde{N}_p\tilde{\zeta}_{G}^p+\frac{1}{(p+1)!}\tilde{N}_{p+1}\tilde{\zeta}_{G}^{p+1}+\dots\;,\;\;p>1.
\ee
where again we assume for simplicity that the first term in the series dominates. In this case the moments $\tilde{\mathcal{M}}_n$ are all of $O(1)$. In the smaller volume $M$ the entire local ansatz series is regenerated, but with
\be
N_n\approx\frac{(1+\tilde{\zeta}_{l,M}(B))^{n-1}}{(\tilde{N}_p(\tilde\sigma_0B)^p)^{n-1}}\frac{((p-1)!)^n}{(p-n)!},\ \ \ n\leq p
\ee
Now the linear term is regenerated like all the other terms, and the correlation functions $\langle\zeta^n\rangle$ are of order $N_2^{n-2}$,
\be
\mathcal{M}_n\propto[f_{NL}^{eff}\sigma_0]^{n-2}\;,\;\;2<n\leq p
\ee
where $f_{NL}^{eff}=\frac{1+\tilde{\zeta}_{l,M}(B)}{2\tilde{N}_p(\tilde\sigma_0B)^p}$. Although there is no longer an additional small factor suppressing the moments, as in Eq. \eqref{FTLmoments}, the scaling of the moments is otherwise the same as described above (for $n>p$, the moments again fall off with the original scale $\tilde{r}$). For biased enough subsamples the moments can be small and fall off rapidly; even a strongly non-Gaussian model in the large volume generates subsamples that are weakly non-Gaussian.

An easy way to see that Gaussian statistics are recovered on small scales is to consider the simple case $\tilde{\zeta}=\tilde{\zeta}_{G}^2$. Breaking $\tilde{\zeta}_{G}$ into long and short wavelength modes, we have $\tilde{\zeta}=\tilde{\zeta}_{Gl}^2+2\tilde{\zeta}_{Gl}\tilde{\zeta}_{Gs}+\tilde{\zeta}_{Gs}^2$. If the number of background modes is much greater than the number of short-wavelength modes, $\ln(L/M)\gg\ln(M/R)$, then as long as $\tilde{\zeta}_{Gl,M}\sim\langle\tilde{\zeta}_{Gl}^2\rangle^{1/2}$, the linear term will be much larger than the quadratic term. In general, when the scale of the subsamples is small enough, typical subsamples will be sufficiently biased to regenerate the familiar local ansatz. (In the case of a scale-dependent power spectrum where longer wavelength modes have greater power $(n_s<1)$, the bias $B\equiv\tilde{\zeta}_{Gl,M}/\tilde\sigma_0$ from the background increases more rapidly as the subsample size $M$ is decreased, causing the linear term to be boosted in size and the field $\zeta$ to be more Gaussian.)

\textit{Behavior of n-Point Function Shapes.} In specifying $\tilde{\zeta}$, we determine shapes for the $n$-point functions on all scales. In subsamples, these shapes are still present, but (as in the two examples considered here) can be dominated by soft limits from higher $n$-point functions induced by the background. One might think that arbitrarily non-linear terms in $\tilde{\zeta}$ could give arbitrary $k$-dependence to the $n$-point functions. Then the usual local-shape $n$-point functions could be recovered in sufficiently biased small subsamples from very different shapes in the large volume. In the highly non-Gaussian case, Eq. \eqref{HNG}, the $n$-point functions may involve many loops (momentum space integrals), whereas in small subsamples the lower order terms allow the dominant shape to come from tree diagrams. Even for the fine-tuned nearly Gaussian case, Eq. \eqref{WNG}, it is possible for $n$-point functions to be dominated by contributions with many loop integrals, if we remove the earlier requirement that higher order terms fall off by $\tilde{r}\ll1$.

To address this possibility, let us consider the $p$-loop contributions to the two-point function from a given higher order term in the series: $\langle(\tilde{N}_{p+1}\tilde{\zeta}_G^{p+1})_{\mathbf{k_1}}(\tilde{N}_{p+1}\tilde{\zeta}_G^{p+1})_{\mathbf{k_2}}\rangle\in\langle\tilde{\zeta}_{\mathbf{k_1}}\tilde{\zeta}_{\mathbf{k_2}}\rangle$. 
This contribution can be expressed in the form
\be
\tilde{P}_{\zeta}^{p\text{-loop}}\propto\int\prod_{i=1}^pd^3p_i\frac{1}{|\mathbf{k}-\mathbf{p}_p|^3}\left[\prod_{i=1}^{p-1}\frac{1}{|\mathbf{p}_{i+1}-\mathbf{p}_i|^3}\right]\frac{1}{p_1^3}.
\ee
We find that after evaluating $m$ such integrals starting from the right, with a momentum cutoff $L^{-1}$ for all factors in denominators and taking the limit $L^{-1}\ll p_i\ll k_{\text{max}}$, an additional factor of $\ln^m(p_{m+1}L)$ appears, giving $\tilde{P}_{\zeta}^{p\text{-loop}}\propto k^{-3}\ln^p(kL)$. Additional terms are also introduced, but either have weaker momentum dependence or can be discarded in the limit $p_i/k_{\text{max}}\ll1$. The appearance of the scale $L$ in these expressions should not be interpreted as measurability of $L$, since its value is completely degenerate with the amplitude of the power spectrum, the spectral index, and analogous quantities for higher order correlation functions (see, eg, \cite{McDonald:2008sc}).

This analysis can be generalized to the three-point and higher $n$-point functions; an $n$-loop contribution to the bispectrum will involve terms of the form \cite{JasonJoe}
\be
\frac{1}{k_1^3k_2^3}\ln^{m_1}(k_1L)\ln^{m_2}(k_2L)\ln^{m_3}(\text{min}(k_1,k_2)L)+\text{perms.},
\ee
where $\sum m_i=n$ and $m_{1,2,3}$ are the number of loops coming from contractions between different pairs among three terms in the series contributing to the bispectrum. In the squeezed limit, $k_1\rightarrow0$ and $k_2\simeq k_3$, only terms with $m_1=0$ will contribute, so the squeezed limit will still be characterized by the usual $k_1^{-3}$ dependence. We conclude that a local ansatz with arbitrarily fine-tuned coefficients $\tilde{N}_n$ can contribute additional logarithmic $k$-dependence to $n$-point functions, but the behavior in the squeezed limit remains unchanged.

The question of shape is also more complex for higher $n$-point functions in that there are more tree level shapes. For the local model, there are two trispectrum shapes typically discussed: $T_g=g_{NL}P_{G}(k_1)P_{G}(k_2)P_{G}(k_3)$ and $T_{\tau}=\tau_{NL}P_{G}(k_1)P_{G}(k_2)P_{G}(|\mathbf{k}_1+\mathbf{k}_3|)$, with sums over permutations; in our case $\tau_{NL}=(\frac{6}{5}f_{NL})^2$. For the nearly Gaussian ansatz, Eq. \eqref{WNG}, cubic and quadratic terms will be regenerated, with $g_{NL}/f_{NL}^2\sim \tilde{r}^{-(p-1)}\gg1$, so the $T_g$ shape will dominate the $T_{\tau}$ shape in sufficiently biased subsamples.  In the large volume this is also true; the leading term $\langle\tilde{\zeta}_{G,\mathbf{k}_1}\tilde{\zeta}_{G,\mathbf{k}_2}\tilde{\zeta}_{G,\mathbf{k}_3}(\tilde{N}_p\tilde{\zeta}_G^p)_{\mathbf{k}_4}\rangle$ (or $p+1$ for even $p$) has the same momentum dependence.

For the highly non-Gaussian ansatz, Eq. \eqref{HNG}, in sufficiently biased subsamples the quadratic and cubic terms are large and $g_{NL}=O(f_{NL}^2)$ (assuming $p>2$), so the two shapes contribute equally.  In the large volume this is also true because the loop integrals in the trispectrum can be contracted diagrammatically in different ways, contributing terms that approximate both tree level shapes \cite{JasonJoe}. As an exception, for $p=2$ the $\tau_{NL}$ shape dominates in both volumes because the quadratic term is abnormally large compared to the cubic term.

This generalizes to higher $n$-point functions as well: the tree-level shape(s) that are dominant throughout the large volume will also dominate in sufficiently biased small subsamples.  For Eq. \eqref{WNG}, the shape from $\langle\tilde{\zeta}_{G,\mathbf{k}_1}\times...\tilde{\zeta}_{G,\mathbf{k}_{n-1}}(\tilde{N}_{n-1}\tilde{\zeta}^{n-1})_{\mathbf{k}_n}\rangle$ will dominate for any $n$-point function on all scales; for Eq. \eqref{HNG}, contributions to $n$-point functions from $\zeta_G^{m\leq p}$ terms (the regenerated missing terms) will dominate, and the momentum dependence in the large volume will be similar.

\textit{Conclusion.} From these examples we see that for homogeneous and isotropic curvature perturbations in a large volume characterized by an arbitrary set of local terms, one recovers a weakly non-Gaussian series in typical subsamples on sufficiently small scales. This limit, where terms fall off by a characteristic ratio $r$ and the non-Gaussian moments follow a hierarchical scaling $\mathcal{M}_{n+1}/\mathcal{M}_n\sim r$, is therefore statistically natural. (Any model for local non-Gaussianity can be expressed as a superposition of the two specific cases considered here.) Furthermore, the shapes of $n$-point functions cannot change arbitrarily by subsampling. In particular, the characteristic squeezed-limit behavior of the local bispectrum cannot be erased by fine-tuning the coefficients, and is therefore a reliable observational signal of local non-Gaussianity even if the universe is larger than what we observe.

These results suggest two important things for understanding what limits on or detection of non-Gaussianity imply for theories of the primordial universe. First, the form of the local ansatz is protected against changes of scale: although the finiteness of the observable universe means a one-to-one map between observations and theory parameters may not be possible, subsampling does not lead to correlation functions with arbitrary shape in momentum space. Second, these results are independent of a specific dynamical origin for the fluctuations and suggest that purely statistical arguments could be used to define a space of most plausible non-Gaussian models to be tested against observations. Our results are complementary to other statistical restrictions on the relative size of certain moments \cite{Smith:2011if}. Extensions and applications of this result for local, scale-dependent local and non-local non-Gaussianity are in progress \cite{MariUs, JasonJoeUs}.

\textit{Acknowledgements}: We gratefully acknowledge discussions with Niayesh Afshordi, Joe Bramante, Jason Kumar, Louis Leblond and David Seery. We especially thank Marilena LoVerde for early discussions on these issues and for a detailed reading of this report. S.S. is grateful to the Aspen Center for Physics and the NSF Grant \#1066293 for hospitality while some of the ideas for this paper were developed. This work has been supported by the Eberly Research Funds of The Pennsylvania State University. The Institute for Gravitation and the Cosmos is supported by the Eberly College of Science and the Office of the Senior Vice President for Research at the Pennsylvania State University.

%

\end{document}